\documentclass[twocolumn,showpacs,preprintnumbers,amsmath,amssymb]{revtex4}

\usepackage{graphicx}
\usepackage{dcolumn}
\usepackage{bm}

\begin{document}


\title{Topological Quantum Phase Transitions \\ in Topological Superconductors}

\author{M. Cristina Diamantini}
\email{cristina.diamantini@pg.infn.it}
\affiliation{%
INFN and Dipartimento di Fisica, University of Perugia, via A. Pascoli, I-06100 Perugia, Italy
}%

\author{Pasquale Sodano}
\email{pasquale.sodano@pg.infn.it}
\affiliation{%
INFN and Dipartimento di Fisica, University of Perugia, via A. Pascoli, I-06100 Perugia, Italy
}%

\author{Carlo A. Trugenberger}
\email{ca.trugenberger@bluewin.ch}
\affiliation{%
 SwissScientifique, ch. Diodati 10 CH-1223 Cologny, Switzerland} %


\date{\today}

\begin{abstract}
In this paper we  show that BF topological superconductors (insulators) exibit  phase transitions between different topologically ordered phases characterized by different ground state degeneracy on manifold with non-trivial topology. These phase transitions are induced by the condensation (or lack of) of topological defects. 
We concentrate on the (2+1)-dimensional case where the BF model reduce to a mixed Chern-Simons term and we show that the superconducting phase has a ground state degeneracy $k$ and not $k^2$. When the symmetry is $U(1) \times U(1)$, namely when both gauge fields are compact, this model is not equivalent to the sum of two Chern-Simons term with opposite chirality, even if naively diagonalizable. This is due to the fact that $U(1)$ symmetry requires an ultraviolet regularization that make the diagonalization impossible. This can be clearly seen using a lattice regularization, where the gauge fields become angular variables.
Moreover we will show that the phase in which both gauge fields are compact is not allowed dynamically.

\end{abstract}

\maketitle

Quantum phase transitions describe changes in the entanglement
pattern of the complex-valued quantum ground state wave function.
The universality classes of these macroscopic quantum ground states
define the corresponding quantum orders \cite{Wen1}. When there is a
gap in the spectrum, the quantum ordered ground state is called
topologically ordered \cite{Wen2}; its remarkable hallmark is a
ground state degeneracy depending only on  the topology of the
underlying space.

The best known example of topological order is given by Laughlin's
quantum incompressible fluids \cite{Laughlin2} describing the ground
states responsible for the quantum Hall effect \cite{GirvinPrange}.

In \cite{dst} we proposed a superconductivity mechanism which is based on a topologically ordered 
ground state rather than on the usual Landau mechanism of spontaneous symmetry breaking.
Topologically ordered superconductors have a long-distance hydrodynamic action which 
can be entirely formulated in terms of generalized compact gauge fields, the dominant 
term being the topological BF action.

In this paper we will show that BF models exhibit phase transitions between different topologically ordered phases. These phase transitions are induced by the condensation (or lack of) of topological defects. These defects are present due to the compactness of the original gauge model.
Different topological order will be characterized by different ground state degeneracy. 
We will concentrate on the (2+1)-dimensional case where the BF model reduce to a mixed Chern-Simons term. We want to point out that when the symmetry is $U(1) \times U(1)$, namely when the to gauge fields are compact, this model is not equivalent to the sum of two Chern-Simons term with opposite chirality, even if naively diagonalizable. This is due to the fact that $U(1)$ symmetry requires an ultraviolet regularization. This can be typically treated by obtaining the $U(1)$ group from spontaneous symmetry breaking of a non-Abelian group or by formulating the gauge theory on a lattice, in which case, the gauge fields are angular variables. In both regularization methods the ultraviolet cutoffs actually prevents the naive diagonalization of the mixed Chern-Simons term, in one case because of the non-quadratic character of the action and in the other because of the angular character of both gauge fields.
The diagonalization is, instead, permitted when only one of the gauge field is compact or when both are non-compact, i.e. for gauge groups $U(1) \times R$,  $R \times U(1)$ or $R \times R$. As we have shown in \cite{jja}  and confirmed in the proper continuum formulation \cite{fercar} these are actually the only cases that can occur due to the condensation of topological defects, the $U(1) \times U(1)$ case being dynamically excluded. The three corresponding phases are superconducting, superinsulator and metallic phases.

BF theories are  topological theories that can be defined on manifolds $M_{d+1}$ of any 
dimension (here d is the number of spatial dimensions) and play a crucial role in models of two-dimensional gravity
\cite{marito}.
In \cite{jja} we have shown that the BF term also plays a crucial role in the physics of Josephson junction arrays.

The BF term \cite{birmi} is the wedge product of a p-form B  and the curvature $d A$ of a (d-p) form A:

\begin{equation}
S_{BF} = {k \over 2 \pi} \int_{M_{d+1}}  B_p \wedge d A_{d-p}\ ,
\label{abf}
\end{equation}
where $k$ is a dimensionless coupling constant.
This action has a generalized Abelian gauge symmetry under the transformation 
\begin{equation}
B \rightarrow B + \eta \quad , A \rightarrow A + \xi \ ,
\nonumber
\end{equation}
where $\eta$ and $\xi$ are a closed p and a closed (d-p) form, respectively.

The degeneracy of the ground state of the BF theory on a manifold with non-trivial topology  was proven in \cite{gord}.
Consider the model (\ref{abf}) with $k = {k_1 \over k_2}$ on a manifold $M_d \times R_1$, 
with $M_d$ a compact, path-connected , orientable d-dimensional manifold without boundaries. The degeneracy of the ground state is expressed
in terms of the intersection matrix $M_{mn}$ \cite{bott} with $m,n = 1....N_p$ and $N_p$  the rank of the matrix, 
between  p-cycles and  (d-p)-cycles. $N_p$ corresponds to the number of generators of the two homology groups
$H_p(M_d)$ and $H_{d-p}(M_d)$ and is essentially the number of non-trivial cycles on 
the manifold $M_d$. The degeneracy of the ground state is given by $|k_1 k_2 M|^{N_p}$,
where $M$ is the integer-valued determinant of the linking matrix. In our case p = (d-1) and the degeneracy reduces to

\begin{equation}
|k_1 \times k_2 M|^{N_{d-1}} \ .
\label{deged}
\end{equation} 

In this paper we consider the special case of (2+1) dimensions (d=2). In this case also
$B$ becomes a 1-form and, correspondingly the BF term reduces to a
mixed Chern-Simons term with action
\begin{equation}
S_{CS} = {k \over 2 \pi} \int_{M_{2+1}}  A_1 \wedge d B_1\ .
\label{smcs}
\end{equation}
In this case the degeneracy on a manifold of genus $g$ is $( k_1 \times k_2 )^{2g}$, and thus $( k_1 \times k_2 )^{2}$ on a torus

In the application to superconductivity, the conserved current $j_1 = * dB_1$ 
represents the charge fluctuations, while the current 
$J_1 = * dA_1$ describes the conserved fluctuations of vortices. As a consequence,
the form $B_1$ must be considered as a pseudo-vector, while $A_1$ is a vector, as
usual. The BF coupling is thus P- and T-invariant. 

The low-energy effective theory of the superconductor can be entirely expressed in terms
of the generalized gauge fields $A_1$ and $B_1$. The dominant term at long distances is
the BF term; the next terms in the derivative expansion of the effective theory are the
kinetic terms for the two gauge fields (for simplicity of 
presentation we shall assume relativistic invariance), giving:

\begin{eqnarray}
S_{TM} &=  \int_{M_{2+1}} { - 1 \over 2 e^2} d A_1 \wedge * d A_1 +  {k \over 2 \pi} A_1 \wedge d B_1 \nonumber \\
&+ {- 1 \over 2 g^2} d B_1\wedge * d B_1\ ,
\label{topmas}
\end{eqnarray}  
where $e^2$ and $g^2$ are coupling constants of dimension $m$.
Note that in (2+1) dimensions the action \ref{topmas}  is invariant under the duality transformation $e \leftrightarrow  g$, $A_1 \leftrightarrow B_1$.
This action, including its non-Abelian generalization with
kinetic terms was first considered in \cite{roman}.

Naively one could diagonalize [\ref{topmas}] by a transformation $A= {1 \over  2} (a+b)$, 
$B=  (a-b)$, giving
\begin{equation}
S_{BF}(d=2) = {k\over 4\pi} \ \int a \wedge da - {k\over 4\pi} \ \int b \wedge db \ .
\label{newa}
\end{equation}
The result is a doubled Chern-Simons model for gauge fields of opposite chirality.  It is
the simplest example of the class of P- and T-invariant topological phases of strongly
correlated (2+1)-dimensional electron systems considered in \cite{doubled}.

The compactness of the gauge fields allows for the presence of topological defects,
both electric and magnetic.The electric (magnetic)  topological defects  couple to the form
$A_1$ ($B_1$) and are string-like objects  described by a singular closed 1-form $Q_1$ ($M_1$). These forms
represent the singular parts of the field strenghts $dA_1$ and $dB_1$, allowed by the compactness of the gauge
symmetries.
The condensation of topological defects  (or lack of) will lead to different topological phases characterized by a different ground state degeneracy on manifold with non-trivial topology.  To show this  we will use an ultraviolet (lattice) regularization.

In \cite{dst} we have shown that the zero-temperature partition function of (2+1)-dimensional BF may be written on the lattice as
fields $A_{\mu}$ (vector) and $B_{\mu}$ (axial vector) as
\begin{eqnarray}
Z &&= \sum_{\{ Q_0 \} \atop \{ M_0 \} } \ 
\int {\cal D}A_{\mu } \int {\cal D}B_{\mu } \ {\rm exp}(-S)\ ,
\nonumber \\
S &&= \int dt \sum_{\bf x} -{ik \over 2\pi} \ A_{\mu }K_{\mu \nu }B_{\nu } \nonumber \\
&&+ i k A_0 Q_0 + i k B_0 M_0\ .
\label{ac}
\end{eqnarray}
 $K_{\mu \nu}$ is the lattice Chern-Simons term \cite{latticecs}, defined by $K_{00} = 0$, $K_{0i} = -\epsilon_{ij}
d_j$, $K_{i0} = S_i \epsilon_{ij} d_j$ and $K_{ij} = -S_i \epsilon_{ij} \partial_0$, in terms of forward (backward) shift and difference operators $S_i$ ($\hat S_i$) and $d_i$ ($\hat d_i$). Its
conjugate $\hat K_{\mu \nu}$ is defined by $\hat K_{00} = 0$, $\hat
K_{0i} = - \hat S_i \epsilon_{ij} \hat d_j$, $\hat K_{i0} =
\epsilon_{ij} \hat d_j$ and $\hat K_{ij} = - \hat S_j \epsilon_{ij}
\partial_0$. The two Chern-Simons kernels $K_{\mu \nu}$ and $\hat
K_{\mu \nu}$ are interchanged upon integration (summation) by parts
on the lattice.
The topological excitations are described by the integer-valued
fields $Q_0$ and $M_0$ and represent unit charges and vortices
rendering the gauge field components $A_0$ and $B_0$ integers via
the Poisson summation formula; their fluctuations determine the
phase diagram \cite{dst}. The lattice spacing $l$ is assumed $l = 1$.

On the lattice the fields $A_{\mu}$  and $B_{\mu}$ are angular variables \cite{polbo} defined on the interval $[- \pi , \pi ]$. 
If we write $A= {1 \over  2} (a+b)$ and $B=  (a-b)$, it is now clear that the  ultraviolet regularization, required by the presence of topological defects, make  impossible to diagonalize naively  the mixed Chern-Simons term.

In the phase in which electric and magnetic topological defects condense  the partition function requires 
a formal sum also over the form $Q_1$ and $M_1$
\begin{eqnarray}
&&Z = \int {\cal D}A {\cal D}B {\cal D}Q {\cal D}M \nonumber \\ 
&&\exp i {k \over 2 \pi}\int_{M_{2+1}} \left( A_1 \wedge d B_1 + A_1 \wedge *Q_1  +  B_1 \wedge *M_1 \right)  \ .
\label{pfbf}
\end{eqnarray}
Topological excitations may be absorbed into the compact gauge
field  ${A^c}_{\mu}$, ${B^c}_{\mu}$.
This case corresponds to the doubled Chern-Simons with both gauge fields compact described by the action [\ref{smcs}], with a ground state degeneracy $(k_1 \times k_2)^2$ on the torus.

In case only one of the two topological defects condenses we have two dual phases \cite{dst}: when the magnetic excitations are dilute  and the charge excitations condense rendering the system a superconductor: vortex confinement amounts
here to the Meissner effect. If the magnetic excitations condense while the charged ones become dilute: the
system exhibits insulating behavior due to vortex superconductivity accompanied by a charge Meissner effect. Since the two phases are dual we will discuss only the charge condensation phase.

First of all let us notice that associated with the confinement of vortices there is a breakdown of the original U(1)
matter symmetry under transformations $A_1 \to A_1 + d\lambda$. To see this let us consider 
the effect of such a transformation on the partition function (\ref{pfbf}) with an
electric condensate. Upon integration by parts, the exponential of the action acquires 
a multiplicative factor
\begin{equation}
{\rm exp} \ i {k_1\over 2\pi k_2} \left( \int_{M_d , t=+\infty} \lambda \wedge * Q_1
-\int_{M_d , t=-\infty} \lambda \wedge * Q_1 \right) \ .
\label{newc}
\end{equation}
Assuming a constant $\lambda $, we see that the only values for which the partition
function remains invariant are 
\begin{equation}
\lambda = 2\pi \ n \ {k_2 \over k_1}\ , \qquad \qquad n=1 \dots k_1 \ ,
\label{newd}
\end{equation}
which shows that the global symmetry is broken from U(1) to $Z_{k_1}$. 
In this phase all charges different from $n k_1 e$ are screened and the Dirac quantization condition imposes that diluted magnetic vortices are quantized in units ${2 \pi \over k_1 e}n$.
Note that this
is not the usual Landau mechanism of spontaneous symmetry breaking. Indeed, there is
no local order parameter and the order is characterized rather by 
the expectation value of non-local, topological operator the 't Hooft loop \cite{dst}.  The important point is that, in this phase, the ground-state degeneracy on the torus is not $k^2$ but rather only $k$, as we will now show.

By rewriting the charge   topological excitations as the
curl of an integer-valued axial field $\varphi_{\mu}$: $Q_0 =
\epsilon_{ij} d_i \varphi_j$, the partition function become:
\begin{eqnarray}
Z_{LE} &&= \sum_{\{ \varphi_i \} } \
\int {\cal D}A_{\mu } \int {\cal D}B_{\mu } \ {\rm exp}\ (-S_{LE})\ ,
\nonumber \\
S_{LE} &&= -i{k\over 2\pi } \int dt \sum_{\bf x} A_0 K_{0i} \left( B_i + \varphi_i \right)
\nonumber \\
&& + A_i K_{i0} B_0 + A_i K_{ij} B_j \ . \label{ai}
\end{eqnarray}
From (\ref{ai}) one sees that the gauge field components $B_i$ are
angular variables due to their invariance under time-independent
integer shifts. Such shifts do not affect the last term in the
action, which contains a time derivative, and may be reabsorbed in
the topological excitations$\varphi _i$, leaving also the first
term of the action invariant. Topological excitations may be absorbed only into the compact gauge
 ${B^c}_{\mu}$ and one may define
\begin{eqnarray}
&&\sum_{\{ \varphi_i \} } \
\int {\cal D}A_{\mu } \int {\cal D}B_{\mu } \ {\rm exp}\ \left( -S_{LE}\left( A_i, B_i, \varphi_i \right) \right) =
\nonumber \\
&&\int {\cal D} A_{\mu}, \int {\cal D}{B^c}_{\mu } \ {\rm exp}\ \left( -S_{LE}\left( A_i, {B^c}_i \right) \right) \ .
\label{aj}
\end{eqnarray}

The canonical quantization of the low energy effective action
$S_{LE}\left( A_i, {B^c}_i \right) $ is carried out by imposing the
Gauss law constraints associated to the two Lagrange multipliers
$A_0$ and $B_0$, $K_{0i} {B^c}_i = 0$ and $\hat K_{0i}  A_i= 0$ and
then enforcing the usual Weyl gauge condition $ A_0 = 0$, ${B^c}_0 =
0$: the canonical commutation relations read  $\left[ A_i ({\bf x}),
B^c_j ({\bf x}+ {\bf i})\right] = {-i 2\pi\over k} \ \epsilon_{ij}$
where ${\bf i}$ denotes a unit lattice vector in direction $i$.

Since now only one of the two gauge fields is compact, we can introduce two pertinently normalized chiral gauge fields defined
by ${A^L}_i \equiv ({1\over 2} A_i + {B^c}_i)$ and ${A^R}_i \equiv 
({1\over 2} A_i-{B^c}_i)$. Following \cite{elsem}, the non-trivial windings around
closed loops  can be encoded
in the two pairs of global variables:
\begin{eqnarray}
q_L &&= \sum_l {A^L}_1 (l,0) \ , \ p_L = -{k\over 2\pi} \ \sum_l {A^L}_2 (0,l) \ ,
\nonumber \\
q_R &&= \sum_l {A^R}_1 (l,0) \ , \  p_R = -{k\over 2\pi} \ \sum_l {A^R}_2 (0,l) \ ,
\label{aab}
\end{eqnarray}
where the sums run over a period $P_l$ ($l=1,2$) of the torus . As a consequence of the commutation relations between $A_i$
and $B_j$ one has that $\left[ q_i, p_j \right] = i\delta_{ij}$,
$i=R,L$.
The generators of large gauge transformations can thus be written as:
\begin{eqnarray}
U_1(n,m) &&= \exp \left[ 2 \pi i (n p_L + m p_R)  \right]  \nonumber \\
U_2(t,l) &&= \exp \left[- i k (t q_L + l q_R)\right]  \ ,
\label{glg}
\end{eqnarray}
and they satisfy  the algebra:
\begin{equation}
U_1(n,m) U_2(t,l) = U_2(t,l) U_1(n,m) \times \exp \left[ i 2 \pi k (n t + m l) \right] \ .
\label{aglg}
\end{equation}
We note that for $k$ an integer the cocycle is always trivial.

Performing the transformation in the integer $n, m,l,t$:
\begin{equation}
n = x + y \  ; \  m = x - y \ ; \ t = z + s \ ; \ l = z -s  ,
\label{trin}
\end{equation}
 we can write 
\begin{eqnarray} &&U_1(n,m) = U_1(x)U_1(y) =  \nonumber \\
&&= \exp \left[ 2 \pi i x  (p_L +  p_R)  \right] \exp \left[ 2 \pi i y (p_L - p_R)  \right]  \nonumber \\
&&U_2(t,l) = U_2(s)U_2(z) = \nonumber \\
&&= \exp \left[ i k s  (q_L +  q_R)  \right] \exp \left[  i k z (q_L - q_R)  \right] 
\label{glg}
\end{eqnarray}
$ U_1(x)$ and $U_1(y)$,  $U_2(s)$ and $U_2(z)$, are two sets of commuting operators corresponding to the form $A^L$ and $A^R$.

Although one may define the winding numbers separately for the two
chiral sectors, there is only one generator of large gauge
tansformations per homology cycle. This is because only the
$U(1)_{axial}$ subgroup is compact; thus, large gauge
transformations in the left sector, ${A^L}_i \to {A^L}_i + d_i
\lambda$ with $\lambda \left( x_i + P_i \right) = \lambda \left( x_i
\right) + 2\pi m_i$ and $m_i \in {\bf Z}$ for $i=1,2$ must be
combined with the corresponding large gauge transformations ${A^R}_i
\to {A^R}_i + d_i \chi$ with $\chi \left( x_i + P_i \right) = \chi
\left( x_i \right) - 2\pi m_i$ in the right sector.

In [\ref{trin}] this implies $x = z = 0$ making thus one of the generators in [\ref{glg}] act trivially on physical states. From here we can already see that the Hilbert space is only  $=k_1 \times k_2$ on the torus. We will however derive this explicitely for this gauge transformation.
The generators of combined large gauge transformations $q_L \to q_L + 2\pi$, $q_R
\to q_R -2\pi$ and $p_L \to p_L - k$, $p_R \to p_R + k$ are
\begin{eqnarray}
U_1(1) &&=  U_1 = {U^L}_1 {U^R}_1 = {\rm exp} \left[ 2\pi i \left( p_L - p_R
\right) \right] \ ,
\nonumber \\
U_2 (1)&&= U_2 = {U^L}_2 {U^R}_2 = {\rm exp} \left[ i k \left( q_L - q_R \right)
\right] \ , \label{aac}
\end{eqnarray}
and satisfy the algebra $U_1 U_2 = U_2 U_1 \exp{[2 \pi i k]}$ .

One is now ready to compute the ground state degeneracy for a
generic rational value of $k$, i. e. $k=k_1/k_2$. On the torus, the
wave function splits into a part depending on the local variables
($x$) and in a part depending on the global variables defined in
(\ref{aab}); the degeneracy of the ground state is entirely
determined by this second part \cite{deg}. In the Schroedinger
representation, the global variables $q$ are realized as generalized
coordinates and the $p$ are the corresponding momenta; thus, the
wave function is a function of the $q$'s only and must carry a
representation of the algebra of large gauge transformations (LGT).
The representations of the algebra of LGTs are specified by two
angles $\theta_1$ and $\theta_2$, which are entirely determined by
the windings around the 2 homology cycles of the torus;
correspondingly, the ground state wave function must also be labeled
by these two angles, $\psi = \psi_{\theta_1 \theta_2}$. Assuming
$U_1 \ \psi_{\theta_1 \theta_2} = \exp{[i \theta_1]}\ \psi_{\theta_1
\theta_2}$ one has
\begin{eqnarray}
U_1 U_2^l \ \psi_{\theta_1 \theta_2} &&= U_2^l U_1 \exp{[2 \pi i k l]}\ \psi_{\theta_1 \theta_2}
\nonumber \\
&&=\exp{[2 \pi i k l + i \theta_1]}\ U_2^l \ \psi_{\theta_1 \theta_2} \ ,
\label{coml}
\end{eqnarray}
and for $l = k_2$ one obtains the same eigenvalue since $\exp{[2 \pi
i (k_1/k_2 )k_2]} = 1$, which leads to a first set of
$k_2$-degenerate independent states $U_2^l \psi_{\theta_1
\theta_2}$, $l=1\dots k_2$. For $l=k_2$, one has $U_2^{k_2}
\psi_{\theta_1 \theta_2} = \exp{[i \theta_2]}\ \psi_{\theta_1
\theta_2}$.

The most general form of the wave function yielding a representation
of LGT is
\begin{equation}
\psi_{\theta_1 \theta_2} = \sum_{n \in {\bf Z} } \exp{[i (n + \theta_1 /2\pi) q]}\ \psi_{\theta_1 \theta_2}(n) \ ,
\label{fog}
\end{equation}
where $q = (q_L -q_R)/2$ is the normalized axial combination of
winding numbers. Using (\ref{fog}) one has
\begin{equation}
U_2^l \psi_{\theta_1 \theta_2} = \sum_{n \in {\bf Z} }
\exp{[i (n +\theta_1 /2\pi) q + i k q l]}\ \psi_{\theta_1 \theta_2}(n)\ ,
\label{gfo}
\end{equation}
which reduces, for $l= k_2$, to
\begin{eqnarray}
&&\sum_{n \in {\bf Z} } \exp{[i q (n + \theta_1/2\pi + k_1)]}\ \psi_{\theta_1 \theta_2}(n) =
\nonumber \\
&&\exp{[i \theta_2 ]}\  \sum_{n \in {\bf Z} } \exp{[i (n + \theta_1
/2\pi )q]}\ \psi_{\theta_1 \theta_2}(n)  \ . \label{gkf}
\end{eqnarray}
Combining this with $U_2^{k_2} \psi_{\theta_1 \theta_2} = \exp{[i
\theta_2]}\ \psi_{\theta_1 \theta_2}$ enables to derive the
quasi-periodicity condition $\psi_{\theta_1 \theta_2}(n) = \exp{[i
\theta_2]} \ \psi_{\theta_1 \theta_2}(n + k_1)$. For each $l$ there
are $k_1$ independent states; thus,  the dimension of the Hilbert
space is $k_1 \times k_2$ for the torus (generically $(k_1\times k_2)^g$ on genus
$g$ Riemann surfaces). 
 
For the superconducting and its dual insulator phase we have a degeneracy that is only $k_1 \times k_2$, while when both gauge fields are compact  the degeneracy
is $(k_1\times k_2)^2$. From the phase structure analysis of BF topological fluids done in \cite{jja,fercar} we see however that the phase in which both topological defects condense is not present, but, instead, we have either electric or magnetic condensation or both are dilute.
Thus the quantum order of the superconducting/insulator  state BF topological fluids
thus in the universality class defined by the low-energy effective
gauge theory
\begin{eqnarray}
S &&= {k\over 4\pi} \int d^3x \ A^L_{\mu} \epsilon^{\mu \nu \alpha} \partial_{\nu} A^L_{\alpha} \nonumber \\
&&- {k\over 4\pi} \int d^3x \ A^R_{\mu} \epsilon^{\mu \nu \alpha} \partial_{\nu} A^R_{\alpha} \ .
\label{an}
\end{eqnarray}
The action (\ref{an}) involves two separate
Chern-Simons terms of opposite chirality. Contrary to previously
considered examples of "doubled Chern-Simons theories"
\cite{doubled}, however, only the axial(vector) subgroup $U(1)_{\rm axial (vector)}$
of the total gauge group $G = U(1)_L \otimes U(1)_R$ is compact; the
diagonal vector group is the non-compact group ${\bf R (L)}$, reflecting
the fact that the charge (vortex) coupled to it is no good quantum number due
to the existence of a superconducting condensate. Thus, the
corresponding topological order is halved: as a consequence BF topological fluids belongs to a universality class which is different than the one
of conventional doubled Chern-Simons theories.

We analyzed so far only the $T=0$ superconducting phase BF topological fluids; by
means of duality one may derive the properties of the corresponding
insulating phase. What about the quantum transition point? To this
end one should observe that a compact $U(1)$ theory always involves
a scale, determining the radius of the gauge group: thus, at the
quantum critical point both gauge groups must decompactify and
long-range quantum fluctuations are described by a continuum gauge
theory with gauge group ${\bf R}_{\rm vector} \otimes {\bf R}_{\rm
axial}$. As a consequence, both charges and vortices must deconfine
at the critical point, implying that, there, the BF topological fluids are- as
expected \cite{jja} - in a metallic phase; this behavior supports
the recent re-proposed  scenario \cite{Laughlin1} that quantum critical points
are generically described by continuum gauge theories with
deconfined degrees of freedom.

\end{document}